\documentclass{article}

\usepackage[final,nonatbib]{neurips_2020}
\usepackage[utf8]{inputenc} 
\usepackage[T1]{fontenc}    
\usepackage{hyperref}       
\usepackage{url}            
\usepackage{booktabs}       
\usepackage{amsfonts}       
\usepackage{nicefrac}       
\usepackage{microtype}      
\usepackage{adjustbox}

\usepackage{amsmath}
\usepackage{hyperref}
\usepackage{graphicx}
\usepackage{bm,pifont}
\newcommand{\cmark}{\ding{51}}
\newcommand{\xmark}{\ding{55}}
\usepackage{wrapfig,booktabs}

\usepackage{xcolor}

\title{Context-aware Self-supervised Learning for Medical Images Using Graph Neural Network}

\author{Li Sun$^{1}$,$^{\ast}$ Ke Yu$^{1}$\thanks{Equal contribution}, and Kayhan Batmanghelich$^1$\\
  University of Pittsburgh, USA\\
  \texttt{\{lis118, key44, kayhan\}@pitt.edu}}

\begin{document}

\maketitle

\begin{abstract}
Although self-supervised learning enables us to bootstrap the training by exploiting unlabeled data, the generic self-supervised methods for natural images do not sufficiently incorporate the \emph{context}. For medical images, a desirable method should be sensitive enough to detect deviation from normal-appearing tissue of each anatomical region; here, anatomy is the context. We introduce a novel approach with two levels of self-supervised representation learning objectives: one on the regional anatomical level and another on the patient-level. We use graph neural networks to incorporate the relationship between different anatomical regions. The structure of the graph is informed by anatomical correspondences between each patient and an anatomical atlas. In addition, the graph representation has the advantage of handling any arbitrarily sized image in full resolution. Experiments on large-scale Computer Tomography (CT) datasets of lung images show that our approach compares favorably to baseline methods that do not account for the context. We use the learned embedding for staging lung tissue abnormalities related to COVID-19.
\end{abstract}

\section{Introduction}
Self-supervised learning has emerged as a powerful way of utilizing unlabelled data and achieved state-of-the-art performance in the computer vision~\cite{wu2018unsupervised, chen2020simple, he2020momentum}. For medical imaging analysis, a large-scale annotated dataset is rarely available, especially for emerging diseases, such as COVID-19. However, there are lots of unlabeled data available. Thus, self-supervised pre-training presents an appealing solution in this domain. There are some existing works that focus on self-supervised methods for learning image-level representations~\cite{chen2019self, taleb2019multimodal,bai2019self}. 
Despite their success, current methods suffer from two challenges: (1) These methods do not account for anatomical context. For example, the learned representation is invariant with respect to body landmarks, which are highly informative for clinicians.
(2) Current methods rely on fix-sized input. The dimensions of raw volumetric medical images can vary across scans due to the differences in subjects' bodies, machine types, and operation protocols. The typical approach for pre-processing natural images is to either resize the image or crop it to the same dimensions because the convolutional neural network (CNN) can only handle fixed dimensional input. However, both approaches can be problematic for medical images. Taking chest CT for example, reshaping voxels in a CT image may cause distortion to the lung~\cite{singla2018subject2vec}, and cropping images may introduce undesired artifacts, such as discounting the lung volume.

\begin{figure}[htp]
\centering
\vspace{-3mm}
    \includegraphics[width = 0.9 \textwidth]{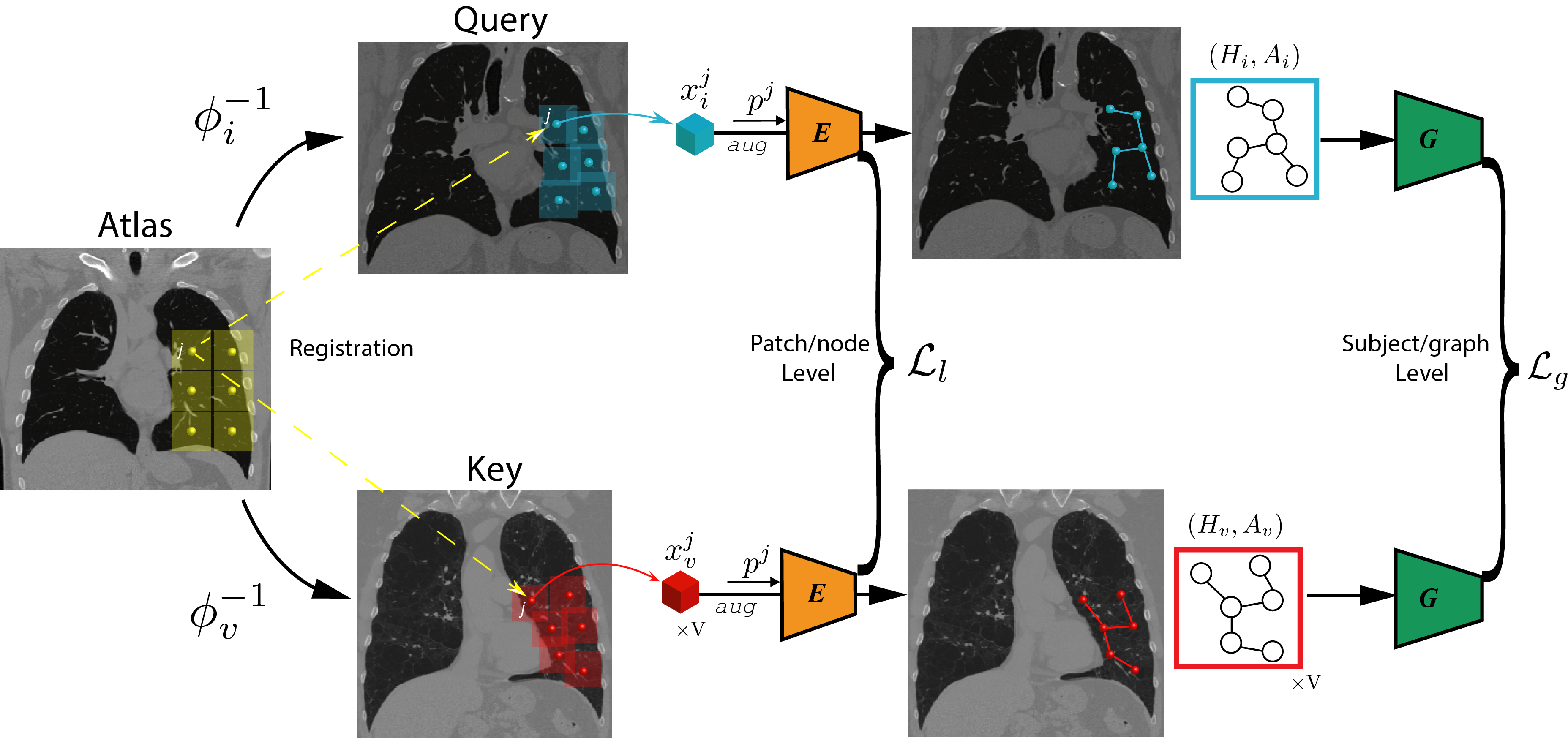}
    \caption{Schematic diagram of the proposed method. 
    We represent every image as a graph of patches. The context is imposed by anatomical correspondences among patients via registration and graph-based hierarchical model used to incorporate the relationship between different anatomical regions. We use a conditional encoder $E(\cdot,\cdot)$ to learn patch-level textural features and use graph convolutional network $G(\cdot,\cdot)$ to learn graph-level representation through contrastive learning objectives.
    }
    \label{fig:main}
\vspace{-5mm}
\end{figure}

To address the challenges discussed above, we propose a novel method for context-aware unsupervised representation learning on volumetric medical images. First, in order to incorporate context information, we represent a 3D image as a graph of patches centered at landmarks defined by an anatomical atlas. The graph structure is informed by anatomical correspondences between the subject's image and the atlas image using registration.
Second, to handle different sized images, we propose a hierarchical model that learns anatomy-specific representations at the patch level and learns subject-specific representations at the graph level. On the patch level, we use a conditional encoder to integrate the local region's texture and the anatomical location.
On the graph level, we use a graph convolutional network (GCN) to incorporate the relationship between different anatomical regions. The schematic is shown in Fig.~\ref{fig:main}.

\section{Method}
\textbf{Constructing Anatomy-aware Graph of Patients} Our method views images of every patient $X_i$ as a set of nodes $\{ x_i^j \}_j^N$, where nodes correspond to image patches covering the lung region of a patient. 
To define a standard set of anatomical regions, we divide the atlas image into a set of $N$ equally spaced 3D patches with some overlap. We register a patient's image to the anatomical atlas by minimizing the following objective function: $Sim(\phi_i(X_i), X_{\text{Atlas}}) + Reg(\phi_i)$, 
where $Sim$ is a similarity metric, $\phi_i(\cdot)$ is the fitted subject-specific transformation, $Reg(\phi_i)$ is a regularization term to ensure the transformation is smooth enough. We use $\{ p^j \}_j^N$ to denote the center coordinates of the patches in the \emph{Atlas coordinate system}. After solving this optimization for each patient, we can use the inverse of this transformation to map $\{ p^j \}_j^N$ to each subject (i.e., $\{ \phi_i^{-1} (p^j) \}$). We use well-established image registration software ANTs~\cite{tustison2014large} to ensure the inverse transformation exists. To incorporate the relationship between different anatomical regions, we represent an image as a graph, whose adjacent matrix $A_i$ is determined by the Euclidean distance between patches' centers and a threshold hyperparameter that controls the density of graph. 

\textbf{Learning Patch-level Representation} We adopt a conditional encoder $E(\cdot,\cdot)$ that takes both patch $x_i^j$ and its location index $j$ as input. It is composed with a CNN feature extractor $C(\cdot)$ and a MLP head $f_l(\cdot)$, thus we have the encoded patch-level feature: $h_i^j=E(x_i^j, j)=f_l(C(x_i^j)\mathbin\Vert p^j)$, where $\mathbin\Vert$ denotes concatenation. We adopt the InfoNCE loss~\cite{oord2018representation} to train the conditional encoder, obtain a positive sample pair by generating two randomly augmented views from the same query patch $x_i^j$,
and obtain a negative sample by augmenting the patch $x_v^j$ at the same anatomical region $j$ from a random subject $v$.

\textbf{Learning Graph-level Representation} We adopt the Graph Convolutional Network (GCN)~\cite{duvenaud2015convolutional} to summarize the patch-level representation into the graph-level representation. Specifically, the GCN model $G(\cdot,\cdot)$ takes patch-level representation $H_i=\texttt{concat}(\{h_i^j\}_{j=1}^N)$ and adjacency matrix $A_i$ as inputs, and propagates information across the graph to update node-level features $H'_i$. We then obtain subject-level representation by global average pooling all nodes in the graph followed by a MLP head $f_g$: $S_i=f_g(\texttt{Pool}(H'_i))$. We adopt the InfoNCE loss to train the GCN on the graph level, obtain a positive pair by taking two views of the same image $X_i$ under random augmentation at the patch level, and obtain a negative sample by randomly sample a different image $X_v$.

\textbf{Overall Model}
The model is trained in an end-to-end fashion by integrating the two InfoNCE losses obtained from patch level and graph level: $\mathcal{L}  =  \mathcal{L}_l(E) +  \mathcal{L}_g(G)$.

\section{Experiments}

\textbf{COPDGene Dataset} We evaluate the proposed method on the COPDGene dataset~\cite{regan2011genetic}, a multi-center observational study designed to identify the underlying genetic factors of the Chronic Obstructive Pulmonary Disease (COPD). We use a large set of 3D CT images of 9,180 subjects from the COPDGene dataset and perform self-supervised pre-training with our method. We then freeze the extracted subject-level features and use them to train linear models to predict various clinically relevant phenotypes (See \textbf{Supplementary Material} for more details). We report the mean $R^2$ scores for regression tasks and the mean accuracies for classification tasks in five-fold cross-validation. 

We compare the performance of our method against: (1) supervised approaches, including  Subject2Vec~\cite{singla2018subject2vec},  Slice-based CNN~\cite{gonzalez2018disease} and (2) unsupervised approaches, including Models Genesis~\cite{zhou2019models}, MedicalNet~\cite{chen2019med3d}, MoCo (3D implementation)~\cite{he2020momentum}, Divergence-based feature extractor~\cite{schabdach2017likelihood}, K-means algorithm applied to image features extracted from local lung regions~\cite{schabdach2017likelihood}, and Low Attenuation Area (LAA), which is a clinical descriptor.

Table~\ref{tbl:COPD} shows that our proposed model outperforms unsupervised baselines
in all metrics except for Future AE. While MoCo is also a contrastive learning based method, we believe that our proposed method achieves better performance for two reasons: (1) Our method incorporates anatomical context. (2) Since MoCo can only accept fixed-size input, we resize all volumetric images into $256\times256\times256$. In this way, lung shapes may be distorted in the CT images, and fine-details are lost due to down-sampling. In comparison, our model supports images with arbitrary sizes in full resolution by design. Our method also outperforms supervised methods, including Subject2Vec and 2D CNN, in terms of CLE, Para-septal, AE History, Future AE, and mMRC; for the rest clinical variables,
the performance gap of our method is smaller than other unsupervised methods. We believe that the improvement is mainly from the richer context information incorporated by our method.
Subject2Vec uses an unordered set-based representation, which does not account for spatial locations of the patches. 2D CNN only uses 2D slices and does not leverage 3D structure. 

\begin{table*}[htp]
\vspace{-5mm}
 \caption{ Evaluation on COPD dataset}
 \begin{adjustbox}{max width=\textwidth}
 \centering
 
 \label{tbl:COPD}
  \begin{tabular}{lc|cc|cccccc}
  \toprule
   Method&Supervised&$\log$\texttt{FEV1pp}&$\log$\texttt{$\text{FEV}_1 / \text{FVC}$}&GOLD&CLE&Para-septal&AE History&Future AE&mMRC\\
   \toprule
   Metric&  &\multicolumn{2}{|c|}{R-Square}& \multicolumn{6}{c}{\% Accuracy}  \\
   \midrule
 LAA-950 &\xmark&$0.44_{\pm.02}$&$0.60_{\pm.01}$&$55.8$&$32.9$&$33.3$&$73.8$&$73.8$&$41.6$\\
 K-Means &\xmark&$0.55_{\pm.03}$&$0.68_{\pm.02}$&$57.3$&-&-&-&-&-\\
 Divergence-based &\xmark& $0.58_{\pm.03}$&$0.70_{\pm.02}$&$58.9$&-&-&-&-&-\\
 MedicalNet &\xmark&$0.47_{\pm.10}$&$0.59_{\pm.06}$&$57.0_{\pm1.3}$&$40.3_{\pm1.9}$&$53.1_{\pm0.7}$&$78.7_{\pm1.3}$&$81.4_{\pm1.7}$&$47.9_{\pm1.2}$\\
 ModelsGenesis&\xmark&$0.58_{\pm.01}$&$0.64_{\pm.01}$&$59.5_{\pm2.3}$&$41.8_{\pm1.4}$&$52.7_{\pm0.5}$&$77.8_{\pm0.8}$&$76.7_{\pm1.2}$&$46.0_{\pm1.2}$\\
 MoCo &\xmark& $0.40_{\pm.02}$&$0.49_{\pm.02}$&$52.7_{\pm1.1}$&$36.5_{\pm0.7}$&$52.5_{\pm1.4}$&$78.6_{\pm0.9}$&\bm{$82.0_{\pm1.2}$}&$46.4_{\pm1.7}$\\
 \midrule
2D CNN &\cmark&$0.53$&-&$51.1$&-&-&$60.4$&-&-\\
Subject2Vec &\cmark&\bm{$0.67_{\pm.03}$}&\bm{$0.74_{\pm.01}$}&\bm{$65.4$}&$40.6$&$52.8$&$76.9$&$68.3$&$43.6$\\
 \midrule
 Ours&\xmark&$0.62_{\pm.01}$&$0.70_{\pm.01}$&$63.2_{\pm1.1}$&\bm{$50.4_{\pm1.3}$}&\bm{$56.2_{\pm1.1}$}&\bm{$78.8_{\pm1.3}$}&$81.1_{\pm1.6}$&\bm{$51.0_{\pm1.0}$}\\
  \bottomrule
   \multicolumn{10}{p{.5\textwidth}}{- indicates not reported.}
  \end{tabular}
   \end{adjustbox}
    \vspace{-3mm}
 \end{table*}

\begin{wraptable}{r}{7cm}
\centering
\caption{ Evaluation on MosMed dataset}
\label{tbl:RU}
\begin{tabular}{lcc}
\toprule
Method&Supervised&\% Accuracy\\
\midrule
MedicalNet &\xmark&$62.1$\\
ModelsGenesis &\xmark&$62.0$\\
MoCo &\xmark&$62.1$\\
3D CNN &\cmark&$61.2$\\
\midrule
Ours&\xmark&\bm{$65.3$}\\
\bottomrule
\end{tabular}
\end{wraptable} 
\textbf{MosMed dataset}
We use 3D CT scans of 1,110 subjects from the MosMed dataset~\cite{morozov2020mosmeddata} provided by municipal hospitals in Moscow, Russia. We first perform self-supervised pre-training with our method on the MosMed dataset. Then we freeze the extracted patient-level features and train a logistic regression classifier to predict the severity of lung tissue abnormalities related to COVID-19, a five-grade categorical variable based on the CT findings and other clinical measures. We compare the proposed method with benchmark unsupervised methods, including MedicalNet, ModelsGenesis, MoCo, and one supervised 3D CNN model. We use the mean test accuracy in five-fold cross-validation as the metric for quantifying prediction performance. Table~\ref{tbl:RU} shows that our proposed model outperforms both the unsupervised and supervised baselines. The supervised 3D CNN model performed worse than the other unsupervised methods, suggesting that it might not converge well or become overfitted since the size of the training set is limited. 

\vspace{-2mm}
\section{Conclusion}
\vspace{-2mm}
We introduce a novel method for context-aware unsupervised representation learning for medical images. Experiments on multiple datasets demonstrate that our method is effective and generalizable.

\section*{Broader Impact}

Our research advances self-supervised representation learning for medical images by exploiting anatomical context embedded in them. For potential practical benefits, our method enables knowledge transformation from existing datasets to detecting emerging diseases, such as COVID-19. In addition, our method doesn't require hand-crafted label, so it avoids selection bias in the setting of supervised learning with limited data. To the best of our knowledge, we are not aware of any impact of our work that has negative ethical and societal consequences.

\bibliographystyle{IEEEtran}
\bibliography{main.bib}

\begin{thebibliography}{10}
\providecommand{\url}[1]{#1}
\csname url@samestyle\endcsname
\providecommand{\newblock}{\relax}
\providecommand{\bibinfo}[2]{#2}
\providecommand{\BIBentrySTDinterwordspacing}{\spaceskip=0pt\relax}
\providecommand{\BIBentryALTinterwordstretchfactor}{4}
\providecommand{\BIBentryALTinterwordspacing}{\spaceskip=\fontdimen2\font plus
\BIBentryALTinterwordstretchfactor\fontdimen3\font minus
  \fontdimen4\font\relax}
\providecommand{\BIBforeignlanguage}[2]{{%
\expandafter\ifx\csname l@#1\endcsname\relax
\typeout{** WARNING: IEEEtran.bst: No hyphenation pattern has been}%
\typeout{** loaded for the language `#1'. Using the pattern for}%
\typeout{** the default language instead.}%
\else
\language=\csname l@#1\endcsname
\fi
#2}}
\providecommand{\BIBdecl}{\relax}
\BIBdecl

\bibitem{wu2018unsupervised}
Z.~Wu, Y.~Xiong, S.~X. Yu, and D.~Lin, ``Unsupervised feature learning via
  non-parametric instance discrimination,'' in \emph{Proceedings of the IEEE
  Conference on Computer Vision and Pattern Recognition}, 2018, pp. 3733--3742.

\bibitem{chen2020simple}
T.~Chen, S.~Kornblith, M.~Norouzi, and G.~Hinton, ``A simple framework for
  contrastive learning of visual representations,'' \emph{arXiv preprint
  arXiv:2002.05709}, 2020.

\bibitem{he2020momentum}
K.~He, H.~Fan, Y.~Wu, S.~Xie, and R.~Girshick, ``Momentum contrast for
  unsupervised visual representation learning,'' in \emph{Proceedings of the
  IEEE/CVF Conference on Computer Vision and Pattern Recognition}, 2020, pp.
  9729--9738.

\bibitem{chen2019self}
L.~Chen, P.~Bentley, K.~Mori, K.~Misawa, M.~Fujiwara, and D.~Rueckert,
  ``Self-supervised learning for medical image analysis using image context
  restoration,'' \emph{Medical image analysis}, vol.~58, p. 101539, 2019.

\bibitem{taleb2019multimodal}
A.~Taleb, C.~Lippert, T.~Klein, and M.~Nabi, ``Multimodal self-supervised
  learning for medical image analysis,'' \emph{arXiv preprint
  arXiv:1912.05396}, 2019.

\bibitem{bai2019self}
W.~Bai, C.~Chen, G.~Tarroni, J.~Duan, F.~Guitton, S.~E. Petersen, Y.~Guo, P.~M.
  Matthews, and D.~Rueckert, ``Self-supervised learning for cardiac mr image
  segmentation by anatomical position prediction,'' in \emph{International
  Conference on Medical Image Computing and Computer-Assisted
  Intervention}.\hskip 1em plus 0.5em minus 0.4em\relax Springer, 2019, pp.
  541--549.

\bibitem{singla2018subject2vec}
S.~Singla, M.~Gong, S.~Ravanbakhsh, F.~Sciurba, B.~Poczos, and K.~N.
  Batmanghelich, ``Subject2vec: generative-discriminative approach from a set
  of image patches to a vector,'' in \emph{International Conference on Medical
  Image Computing and Computer-Assisted Intervention}.\hskip 1em plus 0.5em
  minus 0.4em\relax Springer, 2018, pp. 502--510.

\bibitem{tustison2014large}
N.~J. Tustison, P.~A. Cook, A.~Klein, G.~Song, S.~R. Das, J.~T. Duda, B.~M.
  Kandel, N.~van Strien, J.~R. Stone, J.~C. Gee \emph{et~al.}, ``Large-scale
  evaluation of ants and freesurfer cortical thickness measurements,''
  \emph{Neuroimage}, vol.~99, pp. 166--179, 2014.

\bibitem{oord2018representation}
A.~v.~d. Oord, Y.~Li, and O.~Vinyals, ``Representation learning with
  contrastive predictive coding,'' \emph{arXiv preprint arXiv:1807.03748},
  2018.

\bibitem{duvenaud2015convolutional}
D.~K. Duvenaud, D.~Maclaurin, J.~Iparraguirre, R.~Bombarell, T.~Hirzel,
  A.~Aspuru-Guzik, and R.~P. Adams, ``Convolutional networks on graphs for
  learning molecular fingerprints,'' in \emph{Advances in neural information
  processing systems}, 2015, pp. 2224--2232.

\bibitem{regan2011genetic}
E.~A. Regan, J.~E. Hokanson, J.~R. Murphy, B.~Make, D.~A. Lynch, T.~H. Beaty,
  D.~Curran-Everett, E.~K. Silverman, and J.~D. Crapo, ``Genetic epidemiology
  of copd (copdgene) study design,'' \emph{COPD: Journal of Chronic Obstructive
  Pulmonary Disease}, vol.~7, no.~1, pp. 32--43, 2011.

\bibitem{gonzalez2018disease}
G.~Gonz{\'a}lez, S.~Y. Ash, G.~Vegas-S{\'a}nchez-Ferrero, J.~Onieva~Onieva,
  F.~N. Rahaghi, J.~C. Ross, A.~D{\'\i}az, R.~San Jos{\'e}~Est{\'e}par, and
  G.~R. Washko, ``Disease staging and prognosis in smokers using deep learning
  in chest computed tomography,'' \emph{American journal of respiratory and
  critical care medicine}, vol. 197, no.~2, pp. 193--203, 2018.

\bibitem{zhou2019models}
Z.~Zhou, V.~Sodha, M.~M.~R. Siddiquee, R.~Feng, N.~Tajbakhsh, M.~B. Gotway, and
  J.~Liang, ``Models genesis: Generic autodidactic models for 3d medical image
  analysis,'' in \emph{International Conference on Medical Image Computing and
  Computer-Assisted Intervention}.\hskip 1em plus 0.5em minus 0.4em\relax
  Springer, 2019, pp. 384--393.

\bibitem{chen2019med3d}
S.~Chen, K.~Ma, and Y.~Zheng, ``Med3d: Transfer learning for 3d medical image
  analysis,'' \emph{arXiv preprint arXiv:1904.00625}, 2019.

\bibitem{schabdach2017likelihood}
J.~Schabdach, W.~M. Wells, M.~Cho, and K.~N. Batmanghelich, ``A likelihood-free
  approach for characterizing heterogeneous diseases in large-scale studies,''
  in \emph{International Conference on Information Processing in Medical
  Imaging}.\hskip 1em plus 0.5em minus 0.4em\relax Springer, 2017, pp.
  170--183.

\bibitem{morozov2020mosmeddata}
S.~Morozov, A.~Andreychenko, N.~Pavlov, A.~Vladzymyrskyy, N.~Ledikhova,
  V.~Gombolevskiy, I.~A. Blokhin, P.~Gelezhe, A.~Gonchar, and V.~Y. Chernina,
  ``Mosmeddata: Chest ct scans with covid-19 related findings dataset,''
  \emph{arXiv preprint arXiv:2005.06465}, 2020.

\end{thebibliography}

\end{document}


\maketitle

\section{COPD Phenotypes Shown in Paper}
\begin{itemize}
    \item \texttt{FEV1pp}: percent predicted values of Forced Expiratory Volume in one second
    \item \texttt{$\text{FEV}_1 / \text{FVC}$}: ratio between \texttt{FEV1pp} and Forced vital capacity \texttt{FVC}
    \item GOLD: Global Initiative for Chronic Obstructive Lung Disease (GOLD) score, which is a four-grade categorical value indicating the severity of airflow limitation
    \item CLE: Centrilobular emphysema visual score, which is a six-grade categorical value indicating the severity of emphysema in centrilobular
    \item Para-septal: Paraseptal emphysema visual score, which is a three-grade categorical value indicating the severity of paraseptal emphysema
    \item AE history: Acute Exacerbation history, which is a binary variable indicating whether the subject has experienced at least one exacerbation before enrolling in the study
    \item Future AE: Future Acute Exacerbation, which is a binary variable indicating whether the subject has reported experiencing at least one exacerbation at the 5-year longitudinal follow up
    \item mMRC: Medical Research Council Dyspnea Score, which is a five-grade categorical value indicating dyspnea symptom.
\end{itemize}

\section{Network Archtecture}
In the tables below, we show the detailed architectures of conditional encoder $E(\cdot,\cdot)$, including $C(\cdot)$ and $f_l(\cdot)$, and graph convolutional network $G(\cdot,\cdot)$.
\begin{table}[H]
\centering
\caption{Architecture of the $C$ Network}
\begin{adjustbox}{max width=.5\textwidth}
\label{tbl:arch_c}
\begin{tabular}{lcc}

\toprule
Layer
&Filter size, stride
&Output size$(C,D,H,W)$
\\
\midrule
Input
& - & 1$\times$32$\times$32$\times$32
\\
\midrule
Conv3D
& 3$\times$3$\times$3, 1 & 8$\times$32$\times$32$\times$32
\\
BatchNorm+ELU
& - & 8$\times$32$\times$32$\times$32
\\
Conv3D
& 3$\times$3$\times$3, 2 & 8$\times$16$\times$16$\times$16
\\
BatchNorm+ELU
& - & 8$\times$16$\times$16$\times$16
\\
\midrule
Conv3D
& 3$\times$3$\times$3, 1 & 16$\times$16$\times$16$\times$16
\\
BatchNorm+ELU
& - & 16$\times$16$\times$16$\times$16
\\
Conv3D
& 3$\times$3$\times$3, 1 & 16$\times$16$\times$16$\times$16
\\
BatchNorm+ELU
& - & 16$\times$16$\times$16$\times$16
\\
Conv3D
& 3$\times$3$\times$3, 2 & 16$\times$8$\times$8$\times$8
\\
BatchNorm+ELU
& - & 16$\times$8$\times$8$\times$8
\\
\midrule
Conv3D
& 3$\times$3$\times$3, 1 & 32$\times$8$\times$8$\times$8
\\
BatchNorm+ELU
& - & 32$\times$8$\times$8$\times$8
\\
Conv3D
& 3$\times$3$\times$3, 1 & 32$\times$8$\times$8$\times$8
\\
BatchNorm+ELU
& - & 32$\times$8$\times$8$\times$8
\\
Conv3D
& 3$\times$3$\times$3, 2 & 32$\times$4$\times$4$\times$4
\\
BatchNorm+ELU
& - & 32$\times$4$\times$4$\times$4
\\
\midrule
Conv3D
& 3$\times$3$\times$3, 1 & 64$\times$4$\times$4$\times$4
\\
BatchNorm+ELU
& - & 64$\times$4$\times$4$\times$4
\\
Conv3D
& 3$\times$3$\times$3, 1 & 64$\times$4$\times$4$\times$4
\\
BatchNorm+ELU
& - & 64$\times$4$\times$4$\times$4
\\
Conv3D
& 3$\times$3$\times$3, 2 & 64$\times$2$\times$2$\times$2
\\
BatchNorm+ELU
& - & 64$\times$2$\times$2$\times$2
\\
\midrule
Conv3D
& 3$\times$3$\times$3, 1 & 128$\times$2$\times$2$\times$2
\\
BatchNorm+ELU
& - & 128$\times$2$\times$2$\times$2
\\
Conv3D
& 3$\times$3$\times$3, 2 & 128$\times$1$\times$1$\times$1
\\
BatchNorm+ELU
& - & 128$\times$1$\times$1$\times$1
\\
\midrule
Reshape
& - & 1$\times$128
\\
\bottomrule
\end{tabular}
\end{adjustbox}
\end{table}

\begin{table}[H]
\centering
\caption{Architecture of the $f_l$ Network}
\begin{adjustbox}{max width=.5\textwidth}
\label{tbl:arch_fl}
\begin{tabular}{lcc}

\toprule
Layer
&Filter size, stride
&Output size$(C,F)$
\\
\midrule
Input
& - & 1$\times$128,1$\times$3
\\
Concatenation
& - & 1$\times$131
\\
\midrule
Dense
& - & 1$\times$131
\\
ReLU
& - & 1$\times$131
\\
\midrule
Dense
& - & 1$\times$131
\\
ReLU
& - & 1$\times$131
\\
\midrule
Dense
& - & 1$\times$128
\\
\bottomrule
\end{tabular}
\end{adjustbox}
\end{table}

\begin{table}[H]
\centering
\caption{Architecture of the $G$ Network}
\begin{adjustbox}{max width=.5\textwidth}
\label{tbl:arch_g}
\begin{tabular}{lcc}

\toprule
Layer
&Filter size, stride
&Output size$(C,N,F)$
\\
\midrule
Input
& - & 1$\times$581$\times$128
\\
\midrule
GCNLayer
& - & 1$\times$581$\times$128
\\
BatchNorm+ELU
& - & 1$\times$581$\times$128
\\
AveragePooling
& - & 1$\times$1$\times$128
\\
\midrule
Dense
& - & 1$\times$1$\times$128
\\
ReLU
& - & 1$\times$1$\times$128
\\
\midrule
Dense
& - & 1$\times$1$\times$128
\\
ReLU
& - & 1$\times$1$\times$128
\\
\midrule
Dense
& - & 1$\times$1$\times$128
\\
\midrule
Reshape
& - & 1$\times$128
\\
\bottomrule
\end{tabular}
\end{adjustbox}
\end{table}

\section{Implementation Details}
The patch size is set as $32\times32\times32$. We train the proposed model for 30 epochs. We set the learning rate to be $3\times10^{-2}$. We also employed momentum $= 0.9$ and weight decay $= 1\times10^{-4}$ in the Adam optimizer. The batch size at patch level and subject level is set as 128 and 16, respectively. We let the representation dimension $F$ be $128$. The lung region is extracted using lungmask~\cite{hofmanninger2020automatic}. Following the practice in MoCo, we maintain a queue of data samples and use a momentum update scheme to increase the number of negative samples in training; as shown in previous work, it can improve performance of downstream task~\cite{he2020momentum}. The number of negative samples during training is set as $4096$. The data augmentation includes random elastic transform, adding random Gaussian noise, and random contrast adjustment. The temperature $\tau$ is chosen to be $0.2$. There are $581$ patches per subject/graph. The experiments are performed on 2 GPUs, each with 16GB memory. Cosine schedule~\cite{chen2020improved} is used to update the learning rate. For MoCo~\cite{he2020momentum}, we implement a 3D encoder to handle the 3D data and train the model on COPDGene and MosMed dataset.
For ModelsGenesis~\cite{zhou2019models}, we train the model on COPDGene and MosMed dataset with the original setting. For MedicalNet~\cite{chen2019med3d}, since it's training requires segmentation mask, we use pretrained weights provided by the authors.

\section{Model Visualization}
To visualize the learned embedding and understand the model's behavior, we use two methods to visualize the model. The first one is embedding visualization,
we use UMAP~\cite{mcinnes2018umap} to visualize the patient-level features extracted on the COPDGene dataset in two dimension.
In Fig~\ref{fig:gold}, we found that subjects with GOLD score of (0,1) and (3,4) are separable under two dimension. But subjects with GOLD score 2 are scattered. It requires further investigation to understanding of embedding pattern of subjects subjects with GOLD score 2. 
In Fig~\ref{fig:gold}, we can find a trend, from lower-left to upper-right, along which we can see increasing GOLD score.

\begin{figure}[t]
\centering
    \includegraphics[width = .49\textwidth]
    {figures/GOLD.png}
    \caption{Embedding of subjects in 2D using UMAP. Each dot represents one subject colored by the GOLD score. We can find a trend, from lower-left to upper-right, along which we can see increasing GOLD score. 
    }
    \label{fig:gold}
\end{figure}

Understanding how the model makes predictions is important to build trust in medical imaging analysis. In this section, we propose a method that provides task-specific explanation for the predicted outcome.
Our method is expanded from the class activation maps (CAM) proposed by~\cite{zhou2016learning}. 
Without loss of generality, we assume a Logistic regression model is fitted for a downstream binary classification task (e.g., the presence or absence of a disease) on the extracted subject-level features $S'_i$. The log-odds of the target variable that $Y_i=1$ is:
%
\begin{equation}
\log \frac{\mathbb{P}(Y_i = 1 | S'_i) }{\mathbb{P}(Y_i = 0 | S'_i) } = \beta+WS'_i=\beta+ \sum_{j=1}^{N}\underbrace{  W h_{i}'^j }_{M_i^j} 
\end{equation}
where $S'_i=\texttt{Pool}(H'_i)$, the MLP head $f_g$ is discarded when extracting features for downstream tasks following the practice in~\cite{chen2020simple,chen2020improved}, $\beta$ and $W$ are the learned logistic regression weights. Then we have $M_i^j=W h_{i}'^j$ as the activation score of the anatomical region $j$ to the target classification. We use a sigmoid function to normalize $\{M_i^j\}_j^N$, and use a heatmap to show the discriminative anatomical regions in the image of subject $i$. In Fig.~\ref{fig:copd}, we apply the explanation method using the target logit of GOLD score = 4 on a GOLD 4 subject in COPDGene dataset. The dark area on the right lung, where lung tissue is severely damaged, received highest activation value. Figure~\ref{fig:covid} (left) shows the axial view of the CT image of a COVID-19 positive patient, and Figure~\ref{fig:covid} (right) shows the corresponding activation map. The anatomical regions received high activation scores overlap with the peripheral ground glass opacities on the CT image, which is a known indicator of COVID-19. This result suggests that our model can highlight the regions that are clinically relevant to the prediction.

\begin{figure}[t]
\centering
    \includegraphics[width = .49\textwidth]
    {figures/copd.png}
    \caption{An axial view of the activation map on a $GOLD$ 4 subject in COPDGene dataset. Brighter color indicates higher relevance to the disease severity. The figure illustrates that high activation region overlaps with dark area on  the  right  lung,  where  lung  tissue  is  damaged.
    }
    \label{fig:copd}
\end{figure}

\begin{figure}[t]
\centering
    \includegraphics[width = .49\textwidth]
    {figures/covid_v2.png}
    \caption{An axial view of the activation heatmap on a COVID-19 positive subject in the COVID-19 CT dataset. Brighter color indicates higher relevance to the disease severity. The figure illustrates that high activation region overlaps with the ground glass opacities.
    }
    \label{fig:covid}
\end{figure}
\bibliography{egbib.bib}